\documentclass[sn-mathphys,Numbered]{sn-jnl}

\usepackage{graphicx}%
\usepackage{multirow}%
\usepackage{amsmath,amssymb,amsfonts}%
\usepackage{amsthm}%
\usepackage{mathrsfs}%
\usepackage[title]{appendix}%
\usepackage{xcolor}%
\usepackage{textcomp}%
\usepackage{manyfoot}%
\usepackage{booktabs}%
\usepackage{algorithm}%
\usepackage{algorithmicx}%
\usepackage{algpseudocode}%
\usepackage{listings}%

\theoremstyle{thmstyleone}%
\theoremstyle{thmstyletwo}%

\theoremstyle{thmstylethree}%

\begin{document}

\title[Quantum scaling atomic superheterodyne receiver]{Quantum scaling atomic superheterodyne receiver}


\author[1,2]{\fnm{Peng} \sur{Zhang}}

\author*[1,2,3]{\fnm{Mingyong} \sur{Jing}}\email{jmy@sxu.edu.cn}
\equalcont{These authors contributed equally to this work.}

\author[1,2]{\fnm{Zheng} \sur{Wang}}

\author[1,2]{\fnm{Yan} \sur{Peng}}

\author[1,2]{\fnm{Shaoxin} \sur{Yuan}}

\author[1,2]{\fnm{Hao} \sur{Zhang}}

\author[1,2]{\fnm{Liantuan} \sur{Xiao}}

\author[1,2]{\fnm{Suotang} \sur{Jia}}

\author[1,2]{\fnm{Linjie} \sur{Zhang}}

\affil*[1]{\orgdiv{State Key Laboratory of Quantum Optics and Quantum Optics Devices, Institute of Laser Spectroscopy}, \orgname{Shanxi University}, \orgaddress{\city{Taiyuan}, \postcode{030006}, \state{Shanxi}, \country{China}}}

\affil[2]{\orgdiv{Collaborative Innovation Center of Extreme Optics}, \orgname{Shanxi University}, \orgaddress{\city{Taiyuan}, \postcode{030006}, \state{Shanxi}, \country{China}}}

\affil[3]{\orgdiv{State Key Laboratory of Precision Measurement Technology and Instruments, Department of Precision Instrument}, \orgname{Tsinghua University}, \orgaddress{\street{Haidian}, \city{Beijing}, \postcode{100084}, \country{China}}}


\abstract{Measurement sensitivity is one of the critical indicators for Rydberg atomic radio receivers. This work quantitatively studies the relationship between the atomic superheterodyne receiver's sensitivity and the number of atoms involved in the measurement. The atom number is changed by adjusting the length of the interaction area. The results show that for the ideal case, the sensitivity of the atomic superheterodyne receiver exhibits a quantum scaling: the amplitude of its output signal is proportional to the atom number, and the amplitude of its read-out noise is proportional to the square root of the atom number. Hence, its sensitivity is inversely proportional to the square root of the atom number. This work also gives a detailed discussion of the properties of transit noise in atomic receivers and the influence of some non-ideal factors on sensitivity scaling. This work is significant in the field of atom-based quantum precision measurements.}




\maketitle

\section{Introduction}\label{sec1}

In recent years, radio wave electric field (RF E-field) measurements based on Rydberg atoms have developed rapidly and have shown extensive applications in fields of RF E-field metrology \cite{holloway2017atom,simons2021rydberg,shi2023near,norrgard2021quantum}, remote sensing \cite{robinson2021determining,sedlacek2013atom}, astronomical detection   \cite{marr2015fundamentals,jiang2020fundamental}, wireless communication \cite{meyer2018digital,meyer2021waveguide,anderson2020atomic,holloway2020multiple,liu2022deep,meyer2023simultaneous} and navigation \cite{hofmann2012global}. The most attractive feature of the atomic sensor is that its theoretical sensitivity can beat its classical counterparts \cite{fan2015atom,fancher2021rydberg}, which have been developed for over a hundred years and have reached a physical bottleneck.

Much work has been done on achieving the atomic receiver's theoretical sensitivity. The first E-field measurement based on Rydberg atoms was proposed in 1999 and achieved a measurement accuracy of 20 $\mu {\rm V/cm}$ \cite{osterwalder1999using}. In 2012, the minimal measurable E-field of 8 $\mu {\rm V/cm}$ was achieved \cite{sedlacek2012microwave}, and in 2016, a sensitivity of 5  $\mu {\rm V/cm/\sqrt{Hz}}$ was achieved  \cite{kumar2017atom}. In 2020, the proposal of the atomic superheterodyne receiver (atomic superhet) made the sensitivity rapidly advance to 55  ${\rm nV/cm/\sqrt{Hz}}$  \cite{jing2020atomic,gordon2019weak,simons2019rydberg}. After that, comparable sensitivity was achieved in the many-body Rydberg atomic system in 2022  \cite{ding2022enhanced}. However, the current atomic receiver's sensitivity is far from its theoretical expectation \cite{fan2015atom,prajapati2023sensitivity}, and there is still much work to be done to improve the sensitivity.

In our previous work \cite{wang2023noise}, we quantitatively studied the relations between atomic superhet's noise and atom number by changing the size of the flat-top laser beams. However, since the change in the beam size will lead to a change in the transit process, thus a univariate experiment can not realize well, make some noise characteristics cannot be clearly explained. At the same time, previous work did not study the relationship between the atomic receiver signal and atom number. This work achieved a better univariate experiment by adjusting the interaction length to change the atom number. The relationship between sensitivity and atom number is finally obtained by comprehensively studying signal and noise, and the effect of some experimental imperfections is also discussed.

\section{Experimental setup}\label{sec2}

\begin{figure}[ht!]
	\includegraphics[width=0.8\columnwidth]{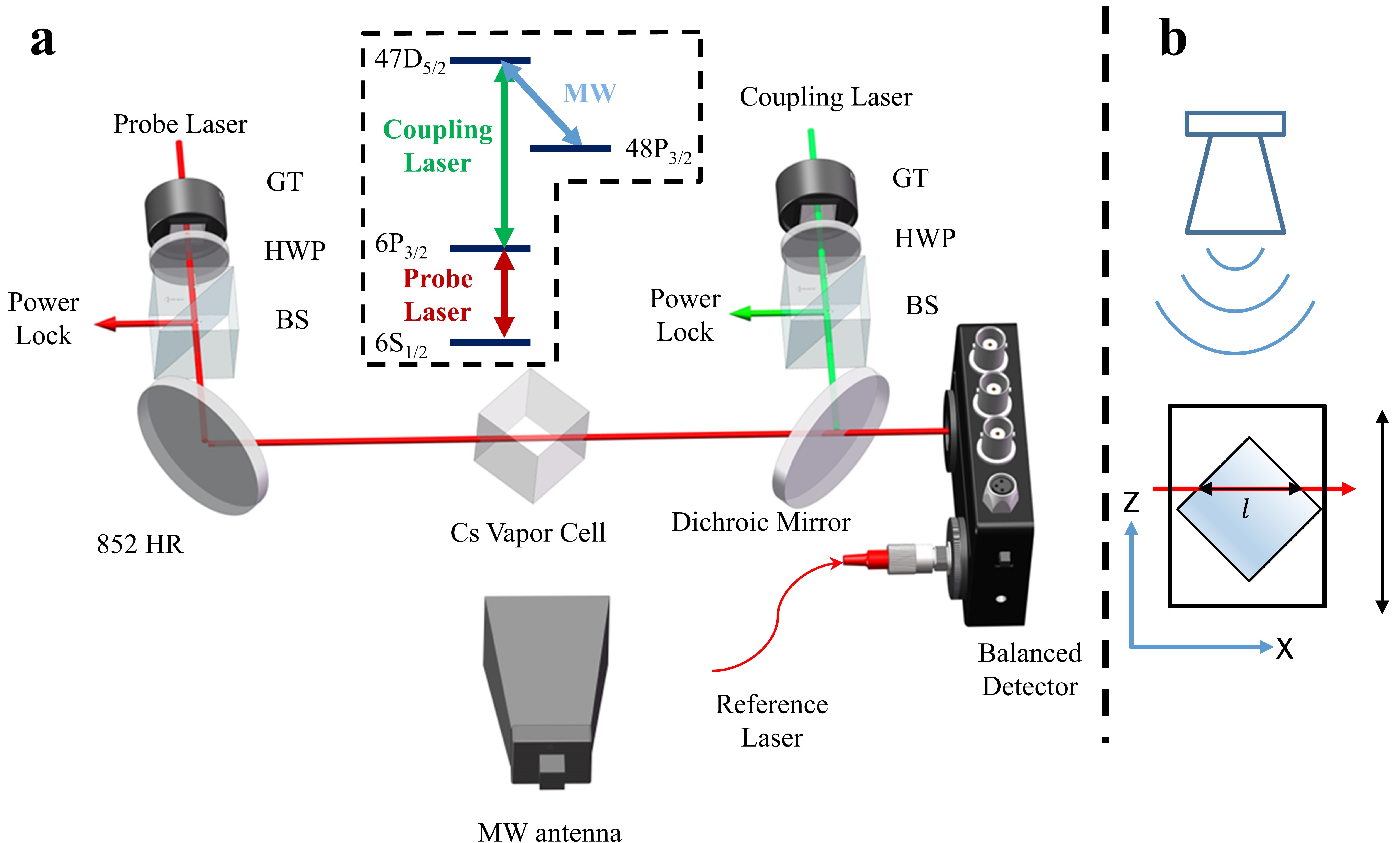}
	\centering
	\caption{\label{fig1}Overall experimental setup and schema. a). Experimental setup. GT: Glan-Taylor Polarizer, HWP: Half-Wave Plate, BS: Beam Splitter (T:R=9:1), and HR: High Reflection dielectric mirror. Inset: Energy level diagram. b). Diagram of changing the interaction length $l$.}
\end{figure}

The primary experimental setup is shown in Fig. \ref{fig1} (a). Inset in Fig. \ref{fig1} (a) is the energy level diagram. We use the same energy level structure as our previous work on atomic superhet \cite{jing2020atomic}. In this work, the $47\rm{D}_{5/2}\rightarrow 48\rm{P}_{3/2}$ transition is coupled only by the local microwave (MW) field when measuring the noise or noise power spectrum (NPS), and is coupled by both signal and local MW field when measuring the signal. The frequency noise of the probe laser and the seed of coupling laser are actively canceled by locking them to a $10$-cm-long ultra-low expansion (ULE) glass cavity with frequency noise servers (FNSs). The FNSs are realized by sideband locking technique whereby the probe and coupling laser frequency can be stabilized to a fixed optical cavity resonance with an adjustable offset  \cite{thorpe2008laser}. The probe and coupling lasers are coupled from polarization-maintaining fibers into free space through collimators.  High-purity vertical polarization of excitation beams is then achieved using Glan Taylor polarizers with more than 60 dB polarization extinction ratio and half-wave plates, which ensures that only $\pi$ transitions can be excited. After polarization purification, probe and coupling lasers are split out 10\% of their power through the unpolarized beam splitters for power monitoring and locking, thereby reducing their intensity noise. The lasers with low frequency and intensity noise help us achieve precision measurements of signal and NPS for atomic superhet. The lasers finally converted into a flat-top profile with a beam radius ($\omega$) of about 1 mm through a beam shaping system (not shown). Eventually, the probe and coupling lasers propagate in the opposite direction and coincide in the cesium vapor cell. Then the probe laser transmission is detected by one of the balanced detector's optical input ports (i.e., signal port). A reference 852 nm laser is input to the other optical port (i.e., reference port) of the balanced detector to cancel the DC offset of photocurrent, which can reduce the photodetector amplifier's saturation and extend the dynamic range of the balanced detector. A spectrum analyzer analyzes the balanced electrical signal outcome by the balanced detector to obtain the power of signal or noise. Signal and local MW fields incident into the atomic vapor cell with vertical polarization and propagate perpendicular to the propagation direction of the probe and coupling laser beams, with Rabi frequencies of about $2\pi \times 7.75$ MHz and $2\pi \times 0.10$ MHz, and frequency of about 6.95 GHz and $6.95+f$ GHz, respectively. Where $f=55$ kHz (read-out frequency) is the frequency difference between the weak signal MW and local MW fields. The multipath propagation effect of microwaves has been suppressed by placing microwave-absorbing materials around the atomic receiver and experiment platform.

This work changes the atom number by adjusting the light-atom interaction length, illustrated in Fig. \ref{fig1} (b). The vapor cell is placed on a linear motorized stage and at an angle of 45 degrees relative to the direction of light propagation. The vapor cell has an inner dimension of $2\times 2\times 2$ mm$^3$. Brushless DC linear servo motor actuators drive the motorized stage and can achieve a high accuracy and repeatability with an absolute on-axis accuracy of $\pm 12.0$ $\mu$m moving along the Z-axis. As the density of the atom number maintains constant, the number of atoms $N_{\rm a}$ that interact with light has a linear relationship with interaction length $l$, i.e., $N_{\rm a}\propto l$. Different from adjusting the atom number by changing the beam radius, which will also change the transit time of the atom moving across the light beam  \cite{wang2023noise}, this method will not change the dynamic response of the atomic receiver. However, this method will change the optical depth of the atomic vapor sample, thus the probe light's absorption rate. In this work, we vary the probe laser's power incident into the vapor cell to maintain its power incident into the balanced detector unchanged (about 30 $\mu$W in our experiment) at different interaction lengths. The Rabi frequencies of the probe laser incident into the vapor cell change from $2\pi \times 7.03$ MHz (16.28 mm length) to $2\pi \times 5.92$ MHz (7.28 mm length). Since the corresponding light intensities of these Rabi frequencies are about four times the saturation intensity, the 20\% variation in Rabi frequency impact on the measurement results can be neglected. The Rabi Frequency of the coupling laser maintains $2\pi \times 0.26$ MHz in the overall experiment.

\section{Experimental results and discussions}
According to the measurement equation of atomic superhet, the relation between the Rabi frequency of signal MW E-field ($\Omega_{\rm s}$) and signal power of optical read-out $P$ at a read-out frequency $f$ is  \cite{jing2020atomic}:
\begin{equation}
	\Omega_{\rm s}(f)=\frac{P(f)}{\kappa_0(f)},\label{measureeq}
\end{equation}
where $\kappa_0(f)$ is the intrinsic conversion gain of the signal MW to the optical read-out of the atomic superhet at read-out frequency $f$. The measurement equation given the measurement sensitivity when optical read-out is only governed by the noise of atomic superhet, i.e., $P(f)=P_{\rm na}(f)$. Thus the sensitivity of atomic superhet is determined by both noise $P_{\rm na}(f)$ and intrinsic conversion gain $\kappa_0(f)$ of atomic superhet.

\begin{figure}[ht!]
\includegraphics[width=\columnwidth]{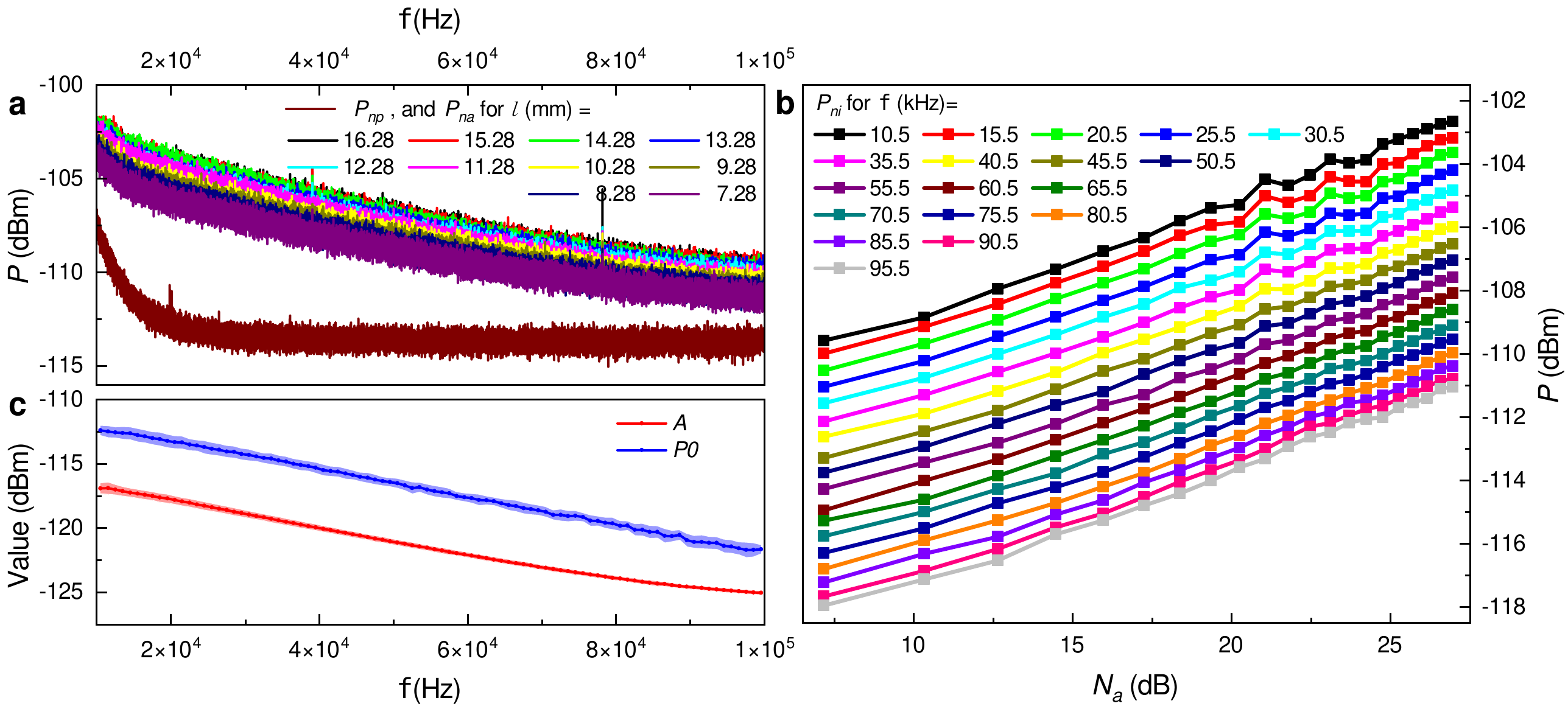}
\centering
\caption{\label{fig2} The noise of atomic superhet at different interaction lengths $l$. a). NPS of probe laser ($P_{\rm np}$) and atomic superhet ($P_{\rm na}$) at different interaction lengths $l$. b). Noise power of interaction noise ($P_{\rm ni}=P_{\rm na}-P_{\rm np}$) as a function of $N_{\rm a}$ at different read-out frequencies $f$, where $N_{\rm a}=20\times\log{l}$ is the relative atom number relative to the number of atoms per unit interaction length ($\rm mm$) in decibel. c). $A$ and $P_0$ (see main text) as a function of read-out frequency. The error bar of data points is the statistical results of five independent experiments.}
\end{figure}

Figure \ref{fig2} shows the noise of atomic superhet at different interaction lengths $l$. In Fig. \ref{fig2} (a), we show the total read-out noise of atomic superhet $P_{\rm na}$ at different $l$ and the intensity noise of the probe laser $P_{\rm np}$ (measured by removing the vapor cell and keeping the power injected to the detector the same). Fig. \ref{fig2} (a) shows the noise of atomic superhet is much larger than the noise of probe laser (at least 5 dB higher); thus, its read-out noise mainly comes from light-atom interaction. The noise of the atomic superhet increases as the interaction length increases. In Fig. \ref{fig2} (b), we subtract the noise of the probe laser from the noise of the atomic superhet to obtain the noise caused by light-atom interaction ($P_{\rm ni}$); then, the data is sectioned at 1 kHz intervals and averaged within each section to reduce the measurement uncertainty. The processed data are regrouped by the same read-out frequency $f$ and rearranged as a function of relative atom number. The data is then fitted by the equation:
\begin{equation}
	P_{\rm ni}=A\times N_{\rm a}^{2\kappa}+P_{\rm n0},
\end{equation}
where $A$ is a proportional coefficient, $P_{\rm n0}$ is a constant term unrelated to the number of atoms, and $\kappa$ is the power-law coefficient of noise amplitude (and $2\kappa$ for noise power) versus the atom number. $P_{\rm n0}$ comes from the residual noise of the measurement system that has not been eliminated from the previous data processing process. The fitting results of $P_0$ decay from -112.40 dBm (at 10 kHz) to -121.64 dBm (at 100 kHz), and is at least 9 dB lower than the noise power of atomic superhet (-103.59 dBm at 10 kHz to -111.07 dBm at 100 kHz), and is a reasonable correction allowed by the experimental error. In this fitting, the power-law coefficient $\kappa$ is kept at 0.5 for best fitting, meaning that the noise amplitude is proportional to the square root of the atom number, which is a typical characteristic of quantum noise. The proportional coefficient $A$ decay as read-out frequency increases and converge to a constant at high frequencies, indicating that although all the interaction noise sources from quantum fluctuations, the quantum fluctuations themselves may originate from different dominated physical processes. In the case of this work, at low read-out frequencies, quantum fluctuations originate from the random transit process of atoms, and at high frequencies, they may come from the random projection of the state vector into one of the states. That is, the interaction noise is dominated by transit noise at low frequencies and may dominated by projection noise at high frequencies. We note that in our previous work  \cite{wang2023noise}, when deducing the power-law coefficient from noise data which is also dominated by transit noise, a power-law coefficient between 0.5 and 1 was obtained, and we reached a misleading conclusion that transit noise is a kind of classical noise, which conflict with results obtained in this work. This is caused by using different methods to change the atom number in these two works, and the details are discussed as follows.

We use a two-level model for intuitive qualitative analysis. For 2-level atoms performing random walks in a resonant-driven Gaussian beam of light, the form of the transit noise power spectrum is  \cite{aoki2016observing}
\begin{align}
	P_{\rm tn}(f)&= \frac{\pi}{4}n_{\rm a} l I_0^2 \omega^2 \sigma_0^2 \int_{-\infty}^\infty \frac{e^{-2 \pi i f \tau}}{1+4 D |\tau|/\omega^2} d \tau \nonumber \\
	&=\frac{I_0^2 \sigma_0^2 N_{\rm a} \phi}{8\pi f}\{-2 \cos{(\phi)}{\rm Ci}(\phi)+\sin{(\phi)}[\pi-2 {\rm Si}(\phi)]\},\label{eq:psd}
\end{align}
where ${\rm Ci}(\phi)$ is the cosine integral function
\begin{equation}
	{\rm Ci}(\phi)=-\int_\phi^\infty d t\cos(t)/t, \nonumber
\end{equation}
${\rm Si}(\phi)$ is the sine integral function
\begin{equation}
	{\rm Si}(\phi)=\int_0^\phi d t \sin(t)/t, \nonumber
\end{equation}
and $\phi=2\pi f \omega^2/4 D$, $D$ is the diffusion constant, $n_{\rm a}$ is the number density of atoms, $\omega$ and $I_0$ is the radius and maximum intensity of Gaussian beam, respectively. $\sigma_0$ is the on-resonance photon absorption cross-section and is given by
\begin{equation}
	\sigma_0=\frac{4\pi\mu^2 f_l}{\hbar c \epsilon_0 \Gamma}, \nonumber
\end{equation}
where $\mu$ is the dipole matrix element, $f_l$ is the linear frequency of resonant-driven light, $\hbar$ is the reduced Planck's constant, $c$ is the speed of light, $\epsilon_0$ is the permittivity of vacuum, and $\Gamma$ is the natural width of transition. $N_{\rm a}=n_{\rm a}\pi \omega^2 l $ is the atom number within interaction region. The square root of asymptotic approximation of eq.\ref{eq:psd} at $f\to 0$, which given the in-band noise amplitude, is
\begin{align}
	\lim_{f\to 0}A_{\rm tn}&=\lim_{f\to 0}\sqrt{P_{\rm tn}}\nonumber\\
	&=\sqrt{\frac{N_{\rm a}\log f}{2 D}}\frac{ I_0 \sigma_0 \omega}{2}\equiv \sqrt{\frac{n_{\rm a} l \log f}{2 D}}\frac{ I_0 \sigma_0 \omega^2}{2} \label{eq:a0},
\end{align}
and the square root of asymptotic approximation of eq.\ref{eq:psd} at $\phi\to\infty$, which given the out-of-band noise amplitude, is
\begin{align}
	\lim_{f\to \infty}A_{\rm tn}&=\lim_{f\to \infty}\sqrt{P_{\rm tn}}\nonumber\\
	&=\sqrt{\frac{D N_{\rm a}}{2}}\frac{I_0 \sigma_0}{\pi f \omega} \equiv \sqrt{\frac{D n_{\rm a} l}{2\pi}}\frac{I_0 \sigma_0}{f}\label{eq:inf}.
\end{align}
Eq. \ref{eq:a0} shows that for in-band noise, its amplitude is proportional to the square root of atom number $\sqrt{N_{\rm a}}$ and the beam radius $\omega$ simultaneously, at the same time $\sqrt{N_{\rm a}}$ itself is also proportional to $\omega$. Therefore its noise amplitude is proportional to $\omega^2$, and a power-law coefficient of 1 to atom number will get if one considers $\omega^2$ and $N_{\rm a}$ equivalent. However, Eq. \ref{eq:a0} shows that only a power-law coefficient value of 0.5 is contributed by varying $N_{\rm a}$ through adjusting beam radius $\omega$, and the other power-law coefficient value of  0.5  is caused by the changing of atomic sensor's instantaneous bandwidth when adjusting the $\omega$. Unlike in-band noise, the amplitude of out-of-band noise (eq. \ref{eq:inf}) does not change relative to beam radius, but it will decrease rapidly with the increase of read-out frequency. At a sufficiently high read-out frequency, it will be well small than the noise contributed by other quantum noise sources.

\begin{figure}[ht!]
	\includegraphics[width=0.8\columnwidth]{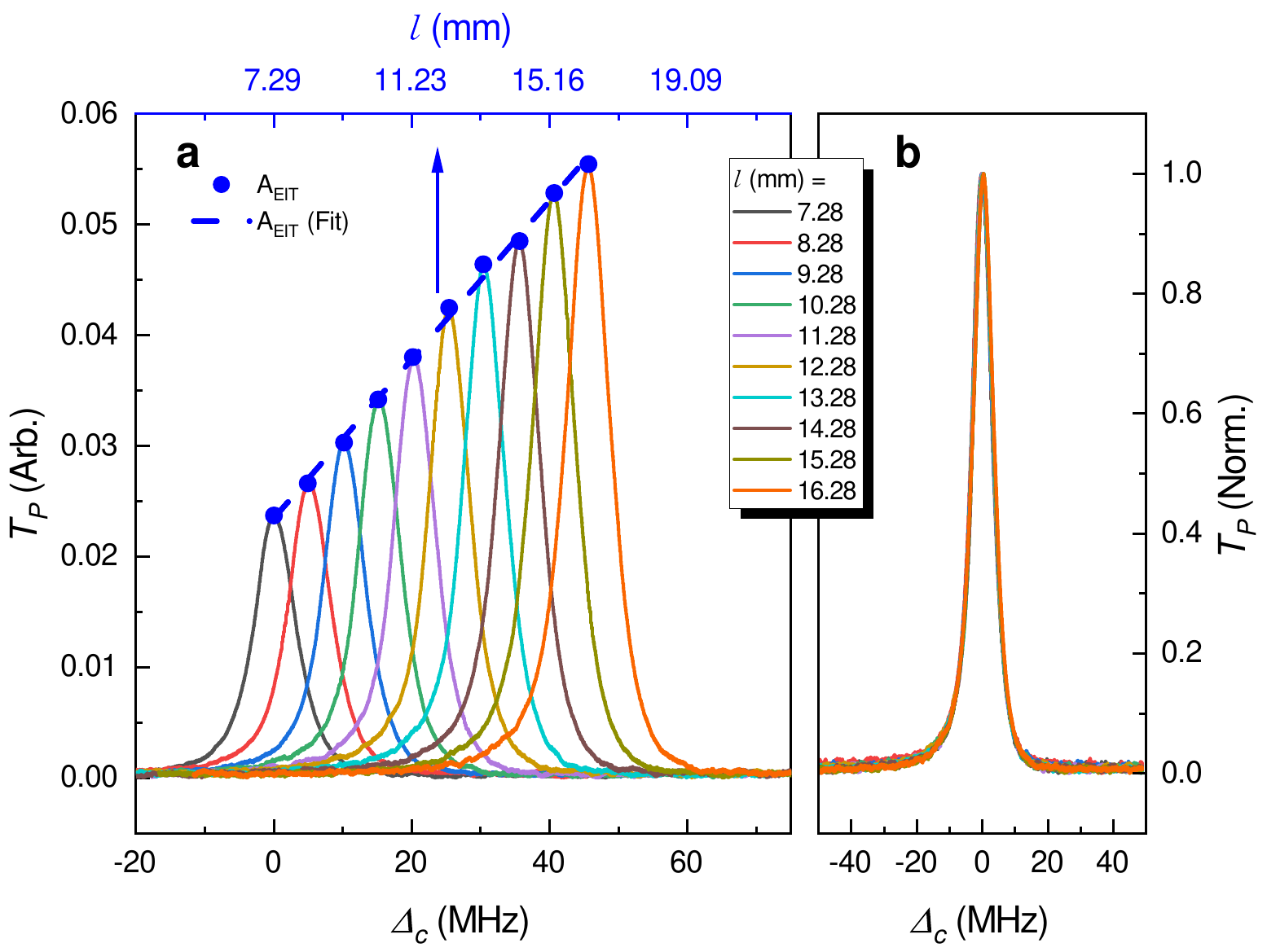}
	\centering
	\caption{\label{fig3} EIT spectra at different interaction lengths $l$. a). Unnormalized EIT spectra at different interaction lengths $l$. Bottom X-axis ($\rm{\Delta_c}$) is the detuning of the coupling laser, and Y-axis ($\rm{T_p}$) is probe transmission. The EIT spectra at different interaction lengths are added with a 5 MHz horizontal shift by hand sequentially to display each spectrum clearly (solid line). A blue dot mark the point of maximum transmission for each spectrum ($A_{\rm EIT}$). The blue dashed line is a linear fit to the maximum transmission as a function of the interaction length (upper X-axis). b). Normalized EIT spectra at different interaction lengths $l$. Y-axis ($\rm{T_p}$) is normalized probe transmission.}
\end{figure}

The intrinsic conversion gain $\kappa_0$ keeps a constant value when the read-out frequency is within the instantaneous bandwidth of atomic superhet, and its value depends on the ratio of amplitude to width of 3-level EIT spectrum  \cite{jing2020atomic}. Figure \ref{fig3} shows the EIT spectra at different interaction lengths (atom number). The unnormalized EIT spectra and linear fit (blue dash line) of EIT amplitude ($A_{\rm EIT}$, blue dot) as a function of interaction length $l$ in Fig. \ref{fig2}(a) shows that the amplitude of EIT spectrum is linear enhance with interaction length $l$. The normalized EIT spectra in Fig. \ref{fig2} (b) show that the full-width-of-half-maximum (FWHM) of the EIT spectrum keeps a constant value of 7.5 MHz when the interaction length changes; thus, for idear case, the intrinsic conversion gain (and the signal amplitude of atomic superhet) increase linear when interaction lengths (atom number) increase. 

\begin{figure}[ht!]
	\includegraphics[width=\columnwidth]{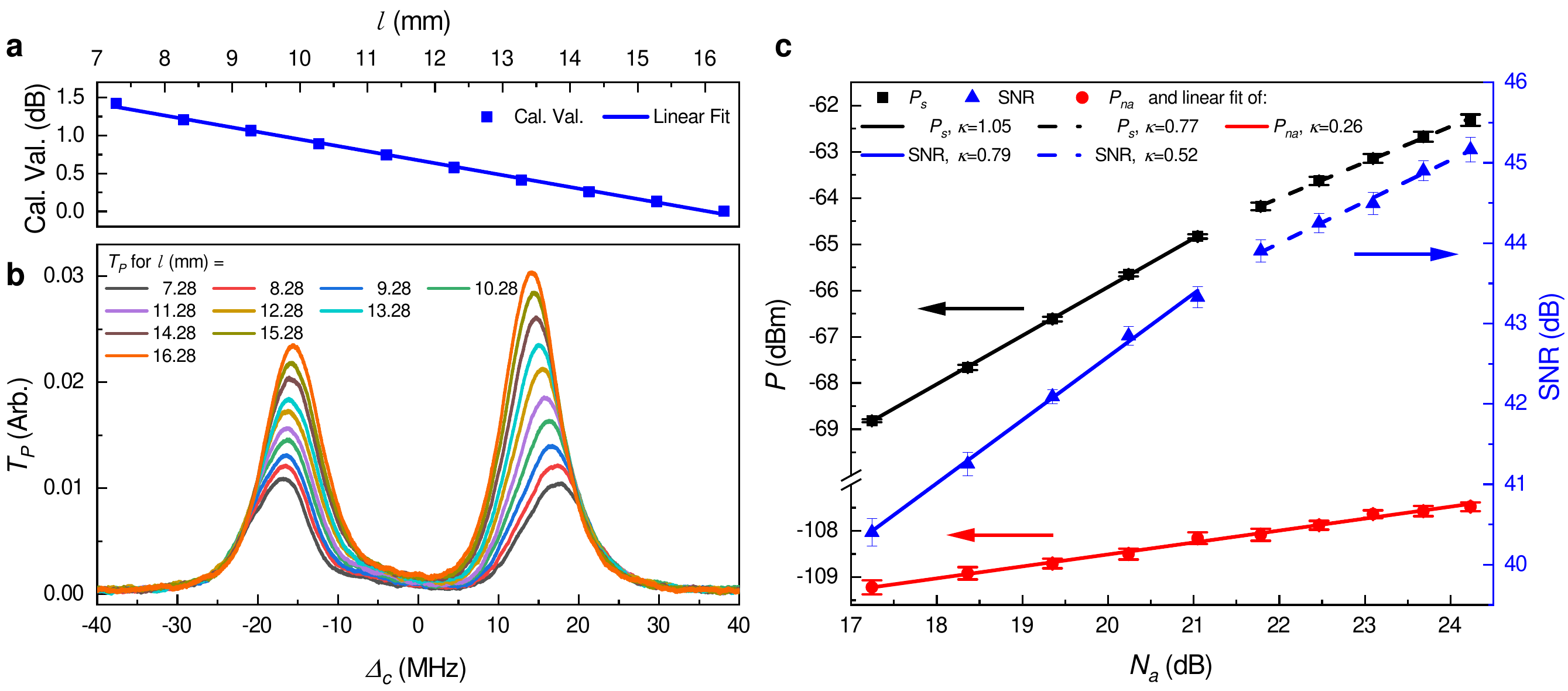}
	\centering
	\caption{\label{fig4} Calibration of MW fields' amplitude, and SNR of atomic superhet at different relative atom numbers $\rm N_a$. a). The calibration value (in dB, blue square dot) of local and signal MW fields as a function of interaction length $l$, and its linear fitting (solid line). b). Autler-Townes (A-T) splitting spectra at different interaction lengths $l$ used to calibrate the MW fields' amplitude. c). Signal power $\rm P_s$ (black square dot), noise power $\rm P_{na}$ (red circle dot), and signal-to-noise ratio (SNR) (blue triangle dot, right Y-axis) of atomic superhet at different relative atom numbers $\rm N_a$. $\rm P_s$, $\rm P_{na}$, and SNR are measured at a read-out frequency of $\delta_s=$ 55 kHz. The solid or dashed straight lines are linear fittings of corresponding data. The error bar of data points is the statistical results of five independent experiments.}
\end{figure}
Since the (power-law) scaling of interaction noise (with atom number) is 0.5, and the scaling of the signal is 1, the theoretical scaling of atomic superhet's sensitivity will be -0.5, obeying a quantum scaling. However, affected by several non-ideal factors, the experimental scaling of atomic superhet's sensitivity is much more complicated. The first non-ideal factor is the microwave field inhomogeneity within the vapor cell, which influences the signal. Figure \ref{fig4} (b) shows that the splitting interval of the AT-splitting spectrum under constant microwave power changes as the interaction length (i.e., the position of the probe and coupling lasers inside the vapor cell) changes and reflects the microwave inhomogeneity inside the vapor cell. According to the splitting interval under different interaction lengths, we make a first-order correction to the microwave power to keep the average E-field amplitude constant at different interaction lengths, and the relationship between the correction value and the interaction length is shown in Fig. \ref{fig4} (a). However, the high-order effect caused by microwave inhomogeneity cannot be eliminated through this correction process, so the experimental scaling of the signal shows a significant difference when interaction length below or over $\lambda/4$ (corresponding to $N_{\rm a}= 21$ dB, and $\lambda$ is the wavelength of the microwave). The black-solid line in Fig. \ref{fig4} (c) shows signal scaling when the interaction length is below $\lambda/4$. In these interaction lengths, the vapor cell cannot support a resonance; thus, the interference inside the cell is significantly reduced, and signal scaling matches the theoretical expectation of 1. The black-dashed line in Fig. \ref{fig4} (c) shows signal scaling when the interaction length is over $\lambda/4$. In these interaction lengths, microwave interference inside the vapor cell induces significant field inhomogeneity and gives a signal scaling of 0.77, significantly smaller than the theoretical expectation. The second non-ideal factor is that the read-out noise consists of more than interaction noise, which influences the noise. The red-solid line in Fig. \ref{fig4} (c) shows a read-out noise scaling of 0.26, significantly minor than the theory expectation of 0.5, and is caused by a constant photon shot noise of probe laser at a level of -113.68 dBm. Consequently, the experimental scaling of SNR is 0.79 when the interaction lengths are below $\lambda/4$ and is 0.52 when the interaction lengths are over $\lambda/4$, around but not strictly equal to the quantum scaling of 0.5.

\section{Conclusions}\label{sec3}
In this work, the number of atoms participating in the light-atom interaction is changed by changing the interaction length, and the power-law scaling of the signal, noise, and signal-to-noise ratio of the atomic superhet with the number of atoms is quantitatively studied. Results show that the atomic superhet's noise is currently dominated by quantum noise, generated by different primary random processes at different read-out frequencies, such as the transit of atoms at low frequency or projection of atomic states at high frequency. For the ideal case, the scaling of the signal-to-noise ratio (sensitivity) of the atom superhet to the number of atoms satisfies the quantum scaling; that is, the SNR (sensitivity) is proportional to $\sqrt{N_{\rm a}}$ (1/$\sqrt{N_{\rm a}}$). However, due to non-ideal factors such as the inhomogeneity of the microwave within the vapor cell and the fact that the read-out noise is not only composed of interaction noise, the experimental scaling of SNR and sensitivity with the number of atoms is close, but not strictly consistent with the quantum scaling, and these are discussed in detail in the paper. This work paves the way for the atomic receiver to improve its experimental sensitivity. E.g., increasing the interaction length to increase the number of atoms is one useful method to improve its sensitivity.

\backmatter


\section*{Declarations}
\begin{itemize}
\item Funding: This research is funded by the National Key R\&D Program of China (grant no. 2022YFA1404003),  the National Natural Science Foundation of China (grants 12104279, 61827824 and 61975104), Innovation Program for Quantum Science and Technology (Grant No. 2021ZD0302100), Shanxi Provincial Key R\&D Program (202102150101001), the Fund for Shanxi ‘1331 Project’ Key Subjects Construction, Bairen Project of Shanxi Province, China, Science and Technology on Electronic Information Control Laboratory Fund.
\item Conflict of interest/Competing interests: The authors declare no conflicts of interest.
\item Availability of data and materials: Data underlying the results presented in this paper are not publicly available at this time, but may be obtained from the authors upon reasonable request.
\item Authors' contributions: M.J. proposed the project, developed the research and analysed the experimental data. P.Z., Z.W. and M.J. performed the experiments. M.J. and Y.P. carried out theoretical calculations. M.J., L.Z. and H.Z. contributed to the experimental set-up. M.J. wrote the manuscript. All authors contributed to discussions of the results and the manuscript and provided revisions of manuscript.
\end{itemize}



\bibliography{sn-bibliography}

\end{document}